\documentclass[letterpaper,12pt]{article}

\usepackage{latexsym}
\usepackage{graphicx}
\usepackage{dcolumn}
\usepackage{amsmath}
\usepackage{amssymb}

\renewcommand{\deg}{^\circ}
\newcommand{\ii}{\mathrm{i}}
\newcommand{\dd}{\mathrm{d}}

\begin{document}

\title{Polarization change induced by a galvanometric optical scanner}

\author{\small \textsc{Gabriele~Anzolin,$^{1,*}$ Arnaud~Gardelein,$^{1}$ Marc~Jofre,$^{1}$} \\
				\small \textsc{Gabriel~Molina-Terriza,$^{1,2}$ and Morgan~W.~Mitchell$^{1}$} \\
				\\
				\small $^1$ICFO-Institut de Ci\`{e}ncies Fot\`{o}niques, Parc Mediterrani de la Tecnologia, \\
				\small 08860 Castelldefels (Barcelona), Spain \\
				\small $^{2}$ICREA-Instituci\'{o} Catalana de Recerca i Estudis Avan\c{c}ats, \\
				\small 08010 Barcelona, Spain \\
				\small $^*$\textit{Corresponding author: gabriele.anzolin@icfo.es}}
				
\date{\small OCIS codes: 000.1600, 120.5700, 200.4860, 260.5430.}

\maketitle

\begin{abstract}
We study the optical properties of a two-axis galvanometric optical scanner constituted by a pair of
rotating planar mirrors, focusing our attention on the transformation induced on the polarization state
of the input beam. We obtain the matrix that defines the transformation of the propagation direction of
the beam and the Jones matrix that defines the transformation of the polarization state. Both matrices
are expressed in terms of the rotation angles of two mirrors. Finally, we calculate the parameters of
the general rotation in the Poincar\'{e} sphere that describes the change of polarization state for each
mutual orientation of the mirrors.
\end{abstract}

\section{Introduction} \label{sec:intro}

An optical scanner is a device constituted by two rotating planar mirrors which are used to deflect a laser beam
along two perpendicular directions \cite{zook74}. Among the different scanning techniques already developed,
galvanometer-based scanners (galvos) offer flexibility, speed and accuracy at a relatively low cost. In fact,
optical scanners based on the current galvo technology permit to obtain closed-loop bandwidths of several kHz and
step response times in the 100~$\mu\mathrm{s}$ range even for beams with large radii. Moreover, a resolution at
the $\mu\mathrm{rad}$ level can be achieved within a large scanning field, which is usually of the order of $20\deg$.

Because of this superb properties, galvo scanners are the preferred solutions in many industrial and scientific
applications requiring fast and precise beam steering capabilities, like medical imaging, information handling,
laser display and material processing \cite{laserscan}. In addition, galvos could find potential applications
in any practical context where the light beam which should be steered and/or stabilized with high precision also
has a well defined state of polarization, like in interferometry with polarized light \cite{interf}, in ellipsometry
\cite{ellips} or in polarization-sensitive optical coherence tomography \cite{optcohtom}. Another important
application might be in single-photon polarization-based quantum communications and quantum key distribution
between two moving terminals, where a galvo scanner placed at the transmitter could be used to point and track the
receiver with high accuracy. However, as the incidence angles of the beam with the two mirrors vary in function of
their mutual position, the corresponding Fresnel coefficients \cite{bornwolf} are subjected to a time-dependent
change that affects the polarization state of the input beam. Therefore, in this kind of applications it is of
primary importance to understand how the polarization state of the output beam changes in function of the combined
motion of the mirrors. 

The problem of the propagation of a polarized beam within a galvo scanner does not seem to have been treated
before in the literature, since previous works were mainly aimed at studying beam path and image distortions
\cite{likatz95,li95,xieetal05,li08}.  Moreover, previous studies of the changes of the polarization state caused
by reflections of a light beam upon moving mirrors refer to optical configurations quite different from that of
an optical scanner, like the Coud\'{e} focus of a telescope \cite{borra76}, sky scanners \cite{garetal80} and
coelostats \cite{becetal05}, or to more general cases of two-mirrors pointing and tracking systems
\cite{bonetal06,bonetal07,bonetal09}. For this reason, the principal purpose of the present work is to give a
theoretical description of the effects on the polarization state of a light beam caused by the motion of the two
mirrors of a galvo scanner.

This paper is structured as follows. In Section~\ref{sec:galvoscanner} we introduce the optical configuration of
a galvo scanner and calculate the matrix that gives the propagation direction of the output beam. In
Section~\ref{sec:polchange} we obtain the Jones matrix of the galvo scanner. In Section~\ref{sec:lossless}
we discuss the polarization change under conditions of lossless mirrors, while in Section~\ref{sec:rotop} we give
a description in terms of rotation operators.

\section{The galvo scanner} \label{sec:galvoscanner}

\subsection{Optical scheme} \label{sec:scheme}

\begin{figure}[p]
\centering
\includegraphics[width=10cm]{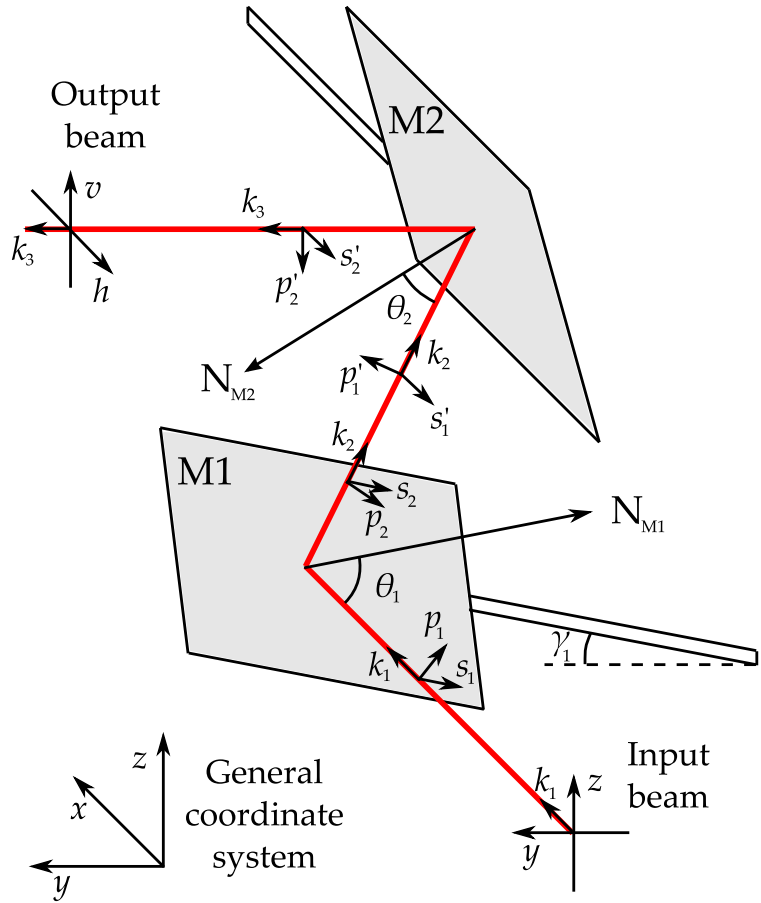}
\caption{\small Schematic view of the galvo scanning system considered in this work, together with the general
coordinate system and the unit vectors used in the calculations. An example of the optical path of a beam is indicated
by the red line.}
\label{fig:galvo}
\end{figure}

In general, a galvo scanner allows to control independently the direction of the output beam along two perpendicular
axis by rotating the two mirrors about the axis of the corresponding galvanometers. We shall call these axis the
``rotation axis'' of the mirrors. The first mirror (M1) controls the deflection of the output beam along the horizontal
direction (left/right), while the second mirror (M2) controls the deflection along the vertical direction (up/down).
Both these deflections are usually referred to a predefined ``zero position'' of M1 and M2. Given a right-handed
reference frame $(x, y, z)$, we consider the galvo mirrors to be at their zero positions when an input beam initially
propagating along the positive $x$ axis will come out along the positive $y$ axis. According to this definition, the wave
vector of the input beam is $\mathbf{k}_1 = [1, 0 ,0]$, while that of the output beam is $\mathbf{k}_3 = [0, 1, 0]$. 

The mutual orientation of the two mirrors when they are at the zero position is a critical parameter that defines the
performances of a galvo scanner. In fact, the distance between their surfaces and the angular range of the scanning
field usually requires the size of M2 to be larger than that of M1. Because of this, M2 is the component that strongly
limits the speed of the entire scanning system.

In the simplest scheme of a galvo scanner, the rotation axis of M1 and M2 correspond with the positive $y$ axis and
the negative $x$ axis, respectively. The zero position is achieved when the vector normal to the reflective surface
of M1 is $[-\sqrt{2}/2, 0, \sqrt{2}/2]$ and the vector normal to the surface of M2 is $[0, \sqrt{2}/2, -\sqrt{2}/2]$.
Starting from this simple configuration, the optical design is usually optimized by rotating M1 by an angle
$\gamma_1 = 15\deg$ about the positive $x$ axis, as schematically shown in Fig.~\ref{fig:galvo}. This configuration
presents the advantage to reduce both width and moment of inertia of M2 and reduce the overall size of the
scanner, with only slight limitations on the speed with respect to the simple scheme. In the following, we will
consider this optimized design because it is the most common configuration of current galvo scanners.

In order to find the zero positions of the mirrors, we have firstly to introduce the criterion to describe rotations
that will be used throughout the paper. Any rotation by an angle $\vartheta$ about an axis $\mathbf{a}$ will be defined
using the ``right-hand rule'', such that a vector $\mathbf{v}$ rotates according to
$\dd \mathbf{v} = \mathbf{a} \times \mathbf{v} \, \dd \vartheta$.  For example, a rotation about the positive $x$ axis
rotates $y$ toward $z$ and $z$ toward $-y$. According to this criterion, the zero position of M1 is obtained by starting
from a configuration in which the normal to its surface is directed along the negative $x$ axis and, then, by rotating
it by $45\deg$ about the positive $y$ axis and by an angle $\gamma_1$ about the positive $x$ axis. It can be easily shown
that the zero position of M2 is simply obtained, starting from a configuration in which the normal to its surface is
directed along the positive $y$ axis, by rotating it about the positive $x$ axis by an angle
$\gamma_2 = 45\deg - \gamma_1/2 = 37.5\deg$. 

The deflection of the output beam along the $x$ axis (horizontal) is the result of a rotation of M1 by an angle $\alpha$
about its rotation axis, starting from its zero position. Therefore, the normal to the surface of M1 is generally defined by
\begin{equation} \label{eqn:norm_m1}
\mathbf{N}_{\mathrm{M1}} =
\left[
\begin{array}{c}
-\cos(\alpha + 45\deg)  \\
-\sin(\alpha + 45\deg) \sin \gamma_1  \\
\sin(\alpha + 45\deg) \cos \gamma_1
\end{array}
\right]~.
\end{equation}
Instead, the motion of the output beam along the $z$ axis (vertical) is achieved by rotating M2 by an angle $\beta$ about
its rotation axis, starting from its zero position. In general, the normal to the surface of M2 has the following vectorial
expression:
\begin{equation} \label{eqn:norm_m2}
\mathbf{N}_{\mathrm{M2}} =
\left[
\begin{array}{c}
0  \\
\cos(\gamma_2 + \beta)\\
-\sin(\gamma_2 + \beta)
\end{array}
\right]~.
\end{equation}

Obviously, when $\alpha \neq 0$ or $\beta \neq 0$, the propagation vector of the output beam, $\mathbf{k}_3$, is no longer
parallel to the $y$ axis.

\subsection{Geometrical transformation matrix} \label{sec:transformation}

Before proceeding with the analysis of the polarization transformation we need to calculate the matrix that transforms
the initial propagation direction $\mathbf{k}_1$ into the final one, $\mathbf{k}_3$. We assume that the mirrors are perfectly
flat, with no deformation occurring while rotating. As a matter of fact, galvo mirrors are usually mass balanced about their
centers of rotation. This solution presents the minimum polar moment of inertia and, therefore, permits to minimize the bending
moments that are generated whenever an eccentric mass is rotated \cite{laserscan}. Under this condition, M1 and M2 have no
dioptric power and the beam can be always considered as collimated throughout its propagation \cite{likatz95}.

The reflection of a ray upon a mirror whose normal has components $[N_x, N_y, N_z]$ is generally described by the matrix
\cite{handopt}
\begin{equation} \label{eqn:refl_mat}
\mathbf{R} = 
\left[
\begin{array}{ccc}
1 - 2 N_x^2  &  -2 N_x N_y  &  -2 N_x N_z  \\
-2 N_x N_y  &  1 - 2 N_y^2  &  -2 N_y N_z  \\
-2 N_x N_z  & -2 N_y N_z  & 1 - 2 N_z^2
\end{array}
\right]~.
\end{equation}
Therefore, the reflection matrix associated to M1 is obtained by using the components of the normal vector shown in
Eq.~\eqref{eqn:norm_m1}:
\begin{equation} \label{eqn:refl_m1}
\footnotesize
\mathbf{R}_{\mathrm{M1}} = 
\left[
\begin{array}{ccc}
\sin(2 \alpha)                & -\cos(2 \alpha) \sin \gamma_1                    & \cos(2 \alpha) \cos \gamma_1  \\
-\cos(2 \alpha) \sin \gamma_1 & \cos^2 \gamma_1 - \sin(2 \alpha) \sin^2 \gamma_1 & [1 + \sin(2 \alpha)] \cos \gamma_1 \sin \gamma_1  \\
\cos(2 \alpha) \cos \gamma_1  & [1 + \sin(2 \alpha)] \cos \gamma_1 \sin \gamma_1 & \sin^2 \gamma_1 - \sin(2 \alpha) \cos^2 \gamma_1
\end{array}
\right]~.
\end{equation}
Instead, the reflection matrix of M2 is obtained by using the vector of Eq.~\eqref{eqn:refl_mat}:
\begin{equation} \label{eqn:refl_m2}
\footnotesize
\mathbf{R}_{\mathrm{M2}} = 
\left[
\begin{array}{ccc}
1  &  0                          &  0  \\
0  &  -\cos[2(\gamma_2 + \beta)] &  \sin[2(\gamma_2 + \beta)]  \\
0  &  \sin[2(\gamma_2 + \beta)]  &  \cos[2(\gamma_2 + \beta)]
\end{array}
\right] =
\left[
\begin{array}{ccc}
1  &  0                        &  0  \\
0  &  -\sin(\gamma_1 + 2\beta) &  \cos(\gamma_1 + 2\beta)  \\
0  &  \cos(\gamma_1 + 2\beta)  &  \sin(\gamma_1 + 2\beta)
\end{array}
\right]~.
\end{equation}

The transformation matrix of the galvo scanning system, i.e. the matrix $\mathbf{G}$ that maps $\mathbf{k}_1$ into
$\mathbf{k}_3$, is finally given by the product between the reflection matrices of the two mirrors. Since any ray of
the input beam is firstly reflected by M1 and then by M2, the $\mathbf{G}$ matrix is given by $\mathbf{G} =
\mathbf{R}_{\mathrm{M2}} \, \mathbf{R}_{\mathrm{M1}}$, or
\begin{equation} \label{eqn:transf_galvo}
\scriptsize
\mathbf{G} = 
\left[
\begin{array}{ccc}
\sin(2 \alpha)  &  -\cos(2 \alpha) \sin \gamma_1
		&  \cos(2 \alpha) \cos \gamma_1  \\
\cos(2 \alpha) \cos(2 \beta)  &  \sin(2 \alpha) \cos(2 \beta) \sin \gamma_1 - \sin(2 \beta) \cos \gamma_1
		&   -\sin(2 \alpha) \cos(2 \beta) \cos \gamma_1 - \sin(2 \beta) \sin \gamma_1  \\
\cos(2 \alpha) \sin(2 \beta)  &  \sin(2 \alpha) \sin(2 \beta) \sin \gamma_1 - \cos(2 \beta) \cos \gamma_1
		&  -\sin(2 \alpha) \sin(2 \beta) \cos \gamma_1 - \cos(2 \beta) \sin \gamma_1
\end{array}
\right]~.
\end{equation}
This matrix clearly illustrates how the rotation of M1 results in a deflection of the output beam along the $x$ axis,
while the rotation of M2 corresponds to a deflection along the $z$ axis. In fact, the final propagation vector of a
beam initially propagating along the positive $x$ axis will be 
\begin{equation}
\mathbf{k}_3 =
\left[
\begin{array}{c}
\sin(2 \alpha)  \\
\cos(2 \alpha) \cos(2 \beta) \\
\cos(2 \alpha) \sin(2 \beta)
\end{array}
\right]
\end{equation}
According to the notation used here, a rotation of M1 by a positive (negative) $\alpha$ angle implies a deflection
towards the positive (negative) $x$, while a rotation of M2 by a positive (negative) $\beta$ is related to a deflection
towards the negative (positive) $z$.

\section{Polarization transformation} \label{sec:polchange}

The calculation of the polarization state of the output beam is made following a procedure quite similar to that explained
in \cite{bonetal06}, which is essentially a polarization ray-tracing approach \cite{chipman95}.

For a galvo scanner, the phenomenon which mainly affects the polarization of the output beam is the reflection upon
the two mirrors. When the characteristics of a mirror are known, the effects on the polarization state of a light beam due
to reflection can be treated using the Jones matrices and the Fresnel coefficients \cite{bornwolf}. These coefficients
give the amount of absorption and phase retard induced by the reflective element on the components of the electric field of
the input beam along the parallel $(p)$ and perpendicular $(s)$ directions to the plane of incidence. Here we consider a
mirror constituted by a single metallic surface with a complex refractive $\tilde{n}$, for simplicity. Assuming propagation
through air, the Fresnel coefficients are defined as
\begin{align}
r_s(\tilde{n}, \theta_i) & = -\frac{\sin(\theta_i - \theta_t)}{\sin(\theta_i + \theta_t)}~, \label{eqn:fresnel_coeff1} \\
r_p(\tilde{n}, \theta_i) & = \frac{\tan(\theta_i - \theta_t)}{\tan(\theta_i + \theta_t)}~, \label{eqn:fresnel_coeff2}
\end{align}
where $\theta_i$ is the incidence angle upon the mirror, $\sin \theta_t = (n_0 \sin \theta_i) / \tilde{n}$ and $n_0$ is
the refractive index of the air. In Eq.~\eqref{eqn:fresnel_coeff1} and \eqref{eqn:fresnel_coeff2} we have neglected the
dependence of the refractive indexes on the wavelength because we are considering a laser beam with a very narrow spectral
bandwidth. 

We consider an input beam propagating along the direction $\mathbf{k}_1 = [1, 0, 0]$. Its polarization plane coincides with
the $(y, z)$ plane and the corresponding Jones vector is $\mathbf{E}_1 = [E_{1y}, E_{1z}]$, where the two components are in
general complex. The beam intersects M1 with an angle of incidence $\theta_1$ given by the dot product
\begin{equation} \label{eqn:deftheta1}
\cos \theta_1 = \mathbf{k}_1 \cdot \mathbf{N}_{\mathrm{M1}}
\end{equation}
and, after being reflected, its direction of propagation is defined by the vector
\begin{equation} \label{eqn:defk2}
\mathbf{k}_2 = \mathbf{R}_{\mathrm{M1}} \mathbf{k}_1~.
\end{equation}

Since the Fresnel coefficients are referred to the $p$ and $s$ directions with respect to the incidence plane,
the Jones vector $\mathbf{E}_1$ has to be expressed in the $(p_1, s_1)$ basis relative the incidence plane with M1:
\begin{equation} \label{eqn:vectors1}
\mathbf{s}_1 = \frac{\mathbf{k}_1 \times \mathbf{k}_2}{\left|\mathbf{k}_1 \times \mathbf{k}_2\right|}~, \;
\mathbf{p}_1 = \frac{\mathbf{s}_1 \times \mathbf{k}_1}{\left|\mathbf{s}_1 \times \mathbf{k}_1\right|}~.
\end{equation}
This change of basis is defined by a 2D rotation of the coordinate system:
\begin{equation} \label{eqn:rot0}
\left[\begin{array}{c}
E_{1p} \\
E_{1s}
\end{array}\right] =
\mathbf{R}(\eta_0)
\left[\begin{array}{c}
E_{1y} \\
E_{1z}
\end{array}\right]~,
\end{equation}
where
\begin{equation}
\mathbf{R}(\vartheta) =
\left[\begin{array}{cc}
\cos \vartheta & \sin \vartheta \\
-\sin \vartheta & \sin \vartheta
\end{array}\right]
\end{equation}
and $\eta_0$ is the angle subtended by $\mathbf{y}$ and $\mathbf{p}_1$.

In general, the $\phi$ angle subtended by two unit vectors $\mathbf{a}$ and $\mathbf{b}$ could be calculated by using
the dot product $\cos \phi = \mathbf{a} \cdot \mathbf{b}$. However, the dot product just provides the smallest positive
angle subtended by the two unit vectors and, therefore, the resulting $\phi$ would be found in the $[0, \pi]$ interval.
To avoid this problem, we introduce the function $\mathrm{ang}(\mathbf{a}, \mathbf{b})$ that provides the $\phi$ angle
in the $[-\pi, \pi]$ range because it can discriminate between clockwise or counterclockwise rotations. In our case,
vectors $\mathbf{a}$ and $\mathbf{b}$ lay in a plane perpendicular to the local propagation vector of the beam,
$\mathbf{k}$. The $\mathrm{ang}$ function is defined as follows: after obtaining the vector
\begin{equation}
\mathbf{c} = \frac{\mathbf{a} \times \mathbf{b}}{\left|\mathbf{a} \times \mathbf{b}\right|}~,
\end{equation}
one has to calculate the scalar $u =  \mathbf{c} \cdot \mathbf{k}$. It turns out that $u = 1$ if the $\mathbf{c}$
and $\mathbf{k}$ vectors are parallel, or $u = -1$ if they are anti-parallel. The correct angle between $\mathbf{a}$
and $\mathbf{b}$ is finally given by
\begin{equation}
\phi = \mathrm{ang}(\mathbf{a}, \mathbf{b}) = u \arccos \left(\mathbf{a} \cdot \mathbf{b}\right)~.
\end{equation}
According to this definition, the rotation angle in Eq.~\eqref{eqn:rot0} is $\eta_0 = \mathrm{ang}(\mathbf{y}, \mathbf{p}_1)$.

The $p$ and $s$ basis associated to the beam reflected by M1 is defined by the vectors
\begin{equation}
\mathbf{s}_2 = \mathbf{s}_1~, \;
\mathbf{p}_2 = \frac{\mathbf{s}_2 \times \mathbf{k}_2}{\left|\mathbf{s}_2 \times \mathbf{k}_2\right|}~,
\end{equation}
The Jones vector $\mathbf{E}_2 = [E_{2p}, E_{2s}]$ of this beam can be calculated using the Jones matrix of M1:
\begin{equation} \label{eqn:jones1}
\left[\begin{array}{c}
E_{2p} \\
E_{2s}
\end{array}\right] =
\left[\begin{array}{cc}
r_{1p}(\tilde{n}_1, \theta_1) & 0 \\
0 & r_{1s}(\tilde{n}_1, \theta_1)
\end{array}\right] \,
\left[\begin{array}{c}
E_{1p} \\
E_{1s}
\end{array}\right]~,
\end{equation}
where $r_{1p}$ and $r_{1s}$ are the complex Fresnel coefficients for M1, while $\tilde{n}_1$ is its complex refractive index.

Then, the beam propagates along $\mathbf{k}_2$ and is reflected by the second mirror. In this case, the incidence angle is
\begin{equation} \label{eqn:deftheta2}
\cos \theta_2 = \mathbf{k}_2 \cdot \mathbf{N}_{\mathrm{M2}}
\end{equation}
and the final direction of propagation is given by
\begin{equation}
\mathbf{k}_3 = \mathbf{R}_{\mathrm{M2}} \, \mathbf{k}_2 = \mathbf{G} \, \mathbf{k}_1~.
\end{equation}
The $p$ and $s$ vectors related to the reflection by M2 are
\begin{equation}
\mathbf{s}'_1 = \frac{\mathbf{k}_2 \times \mathbf{k}_3}{\left|\mathbf{k}_2 \times \mathbf{k}_3\right|}~, \;
\mathbf{p}'_1 = \frac{\mathbf{s}'_1 \times \mathbf{k}_2}{\left|\mathbf{s}'_1 \times \mathbf{k}_2\right|}~,
\end{equation}
for the incident beam, and
\begin{equation}
\mathbf{s}'_2 = \mathbf{s}'_1~, \;
\mathbf{p}'_2 = \frac{\mathbf{s}'_2 \times \mathbf{k}_2}{\left|\mathbf{s}'_2 \times \mathbf{k}_2\right|}~,
\end{equation}
for the reflected beam. While propagating between M1 and M2, the Jones vector of the beam can be expressed in
function of either the $(p_2, s_2)$ basis or the $(p'_1, s'_1)$ basis according to
\begin{equation} \label{eqn:rot1}
\left[\begin{array}{c}
E'_{2p} \\
E'_{2s}
\end{array}\right] =
\mathbf{R}(\eta_1)
\left[\begin{array}{c}
E_{2p} \\
E_{2s}
\end{array}\right]
\end{equation}
where $\eta_1 = \mathrm{ang}(\mathbf{p}_2, \mathbf{p}'_1)$.

Finally, the $p$ and $s$ components of the Jones vector $\mathbf{E}_3$ associated to the output beam are given by
\begin{equation} \label{eqn:jones2}
\left[\begin{array}{c}
E_{3p} \\
E_{3s}
\end{array}\right] =
\left[\begin{array}{cc}
r_{2p}(\tilde{n}_2, \theta_2) & 0 \\
0 & r_{2s}(\tilde{n}_2, \theta_2)
\end{array}\right] \,
\left[\begin{array}{c}
E'_{2p} \\
E'_{2s}
\end{array}\right]~,
\end{equation}
where $r_{2p}$ and $r_{2s}$ are the complex Fresnel coefficients for M2, while $\tilde{n}_2$ is its complex refractive
index. It is useful to project $\mathbf{E}_3$ into a reference system defined by the local horizontal $(h)$ and vertical
$(v)$ directions. The two unit vectors must form a left-handed reference frame together with $\mathbf{k}_3$:
\begin{equation}
\mathbf{h} = \frac{\mathbf{z} \times \mathbf{k}_3}{\left|\mathbf{z} \times \mathbf{k}_3\right|}~, \;
\mathbf{v} = \frac{\mathbf{k}_3 \times \mathbf{h}}{\left|\mathbf{k}_3 \times \mathbf{h}\right|}~.
\end{equation}
Also in this case, the transformation from the $(p'_2, s'_2)$ basis to the $(h,v)$ basis is a 2D rotation:
\begin{equation} \label{eqn:rot2}
\left[\begin{array}{c}
E_{3h} \\
E_{3v}
\end{array}\right] =
\mathbf{R}(\eta_2)
\left[\begin{array}{c}
E_{3p} \\
E_{3s}
\end{array}\right]
\end{equation}
where $\eta_2 = \mathrm{ang}(\mathbf{p}'_2, \mathbf{h})$. All the unit vectors introduced in this Section have been
drawn in Fig.~\ref{fig:galvo} for clarity.

In summary, the Jones matrix describing the transformation of the polarization state of a collimated beam after passing
through the galvo scanner is
\begin{multline} \label{eqn:jones_galvo1}
\mathbf{M} =
\left[\begin{array}{cc}
\cos \eta_2 & \sin \eta_2 \\
-\sin \eta_2 & \cos \eta_2
\end{array}\right]
\left[\begin{array}{cc}
r_{2p} & 0 \\
0 & r_{2s}
\end{array}\right] \times \\
\times \left[\begin{array}{cc}
\cos \eta_1 & \sin \eta_1 \\
-\sin \eta_1 & \cos \eta_1
\end{array}\right]
\left[\begin{array}{cc}
r_{1p} & 0 \\
0 & r_{1s}
\end{array}\right]
\left[\begin{array}{cc}
\cos \eta_0 & \sin \eta_0 \\
-\sin \eta_0 & \cos \eta_0
\end{array}\right]~,
\end{multline}
where we have dropped the explicit dependences on the refractive indexes and the incidence angles, for simplicity.
Since the Fresnel coefficients of the Jones matrices shown in Eqs.~\eqref{eqn:jones1} and \eqref{eqn:jones2} are
in general complex, then both $E_{3h}$ and $E_{2v}$ will be complex and the final polarization state will be in
general elliptical.

\section{Lossless mirrors approximation} \label{sec:lossless}

The Jones matrix of a galvo scanner shown in Eq.~\eqref{eqn:jones_galvo1} is a product of rotation matrices and Jones
matrices of mirrors. The latter have the following common form:
\begin{equation}
\mathbf{A} =
\left[\begin{array}{cc}
r_p & 0 \\
0 & r_s
\end{array}\right] =
\left[\begin{array}{cc}
\rho_p \exp(\ii \phi_p) & 0 \\
0 & \rho_s \exp(\ii \phi_s)
\end{array}\right]
\end{equation}
where we have put in evidence the complex nature of the Fresnel coefficients $r_p$ and $r_s$. The determinant of this
matrix $\det(\mathbf{A}) = \rho_p \rho_s \exp[\ii (\phi_p + \phi_s)]$ is, in general, a complex number different from
unity. For this reason, $\mathbf{A}$ is not unimodular unless $\rho_p \rho_s = 1$ and $\phi_p = -\phi_s$. However,
the $\mathbf{A}$ matrix can be always reduced to a product between a complex constant and an unimodular matrix:
\begin{equation}
\mathbf{A} = \sqrt{\rho_p \rho_s} \, \exp\left(\ii \frac{\phi_p + \phi_s}{2}\right)
\left[\begin{array}{cc}
A \exp(\ii \Phi) & 0 \\
0 & A^{-1} \exp(-\ii \Phi)
\end{array}\right]~,
\end{equation}
where $A = \sqrt{\rho_p / \rho_s}$ and $\Phi = (\phi_p - \phi_s) / 2$.

In the case of mirrors with a high reflectivity it is always found that $0.95 \lesssim \rho_p / \rho_s \lesssim 1$
for a wide range of incidence angles \cite{handopt}. Therefore, the $A$ factor can be approximated to unity (lossless
mirrors) and the Jones matrix becomes totally equivalent to that of a simple phase retarder:
\begin{equation} \label{eqn:mirrorplate}
\mathbf{A} = \sqrt{\rho_p \rho_s} \, \exp\left(\ii \frac{\phi_p + \phi_s}{2}\right)
\left[\begin{array}{cc}
\exp(\ii \Phi) & 0 \\
0 & \exp(-\ii \Phi)
\end{array}\right]~.
\end{equation}
We then assume that the two Jones matrix of the mirrors in Eq.~\eqref{eqn:jones_galvo1} refer to the lossless case.
Under this condition, neglecting the constant phase factor and the attenuation factor in Eq.~\eqref{eqn:mirrorplate},
the approximated version of the Jones matrix of a galvo scanner is given by the product
\begin{multline} \label{eqn:jones_galvo2}
\mathbf{M} \approx
\left[\begin{array}{cc}
\cos \eta_2 & \sin \eta_2 \\
-\sin \eta_2 & \cos \eta_2
\end{array}\right]
\left[\begin{array}{cc}
\exp(\ii \Phi_2) & 0 \\
0 & \exp(-\ii \Phi_2)
\end{array}\right]
\left[\begin{array}{cc}
\cos \eta_1 & \sin \eta_1 \\
-\sin \eta_1 & \cos \eta_1
\end{array}\right] \times \\
\times \left[\begin{array}{cc}
\exp(\ii \Phi_1) & 0 \\
0 & \exp(-\ii \Phi_1)
\end{array}\right]
\left[\begin{array}{cc}
\cos \eta_0 & \sin \eta_0 \\
-\sin \eta_0 & \cos \eta_0
\end{array}\right]~,
\end{multline}
where $\Phi_1 = (\phi_{1p} - \phi_{1s}) / 2$ and $\Phi_2 = (\phi_{2p} - \phi_{2s}) / 2$.

The $\eta_0$ angle is defined as the $\mathrm{ang}$ function of the unit vectors $\mathbf{y}$, which is constant,
and $\mathbf{p}_1$, which is a given by a vector triple product involving only $\mathbf{k}_1$ and $\mathbf{k}_2$.
As shown in Eq.~\eqref{eqn:defk2}, $\mathbf{k}_2$ is the result of the product between the reflection matrix of M1
and $\mathbf{k}_1$, which is constant and coincides with the constant unit vector $\mathbf{x}$ in our assumptions.
Therefore, $\eta_0$ only depends on $\mathbf{R}_{\mathrm{M1}}$, which is a function of the only $\alpha$ and
$\gamma_1$ angles. Following a similar reasoning it can be shown that both $\eta_1$ and $\eta_2$ are functions
of only $\alpha$, $\beta$ and $\gamma_1$.

The phase differences $\Phi_1$ and $\Phi_2$, instead, are related to the complex exponent of the the Fresnel
coefficients, which depends on the complex refractive indexes of the mirrors and the incidence angles. According
to the definitions given in Eq.~\eqref{eqn:deftheta1} and \eqref{eqn:deftheta2}, $\theta_1$ is a function of the constant
vector $\mathbf{k}_1$ and the normal vector to M1, while $\theta_2$ is a function of $\mathbf{k}_2$ and the normal
vector to M2. As a result, we have $\Phi_1 = \Phi_1(\tilde{n}_1, \alpha, \gamma_1)$ and $\Phi_2 = \Phi_2(\tilde{n}_2,
\alpha, \beta, \gamma_1)$.

If the input beam is always kept at a fixed direction and the mechanical properties of the galvo scanner do not change
with time, which means constant values of $\gamma_1$ and of the complex refractive indexes of the mirrors, then the
$\mathbf{M}$ matrix is just a function of the $\alpha$ and $\beta$ angles.

\begin{figure}[p]
\centering
\includegraphics[width=10cm]{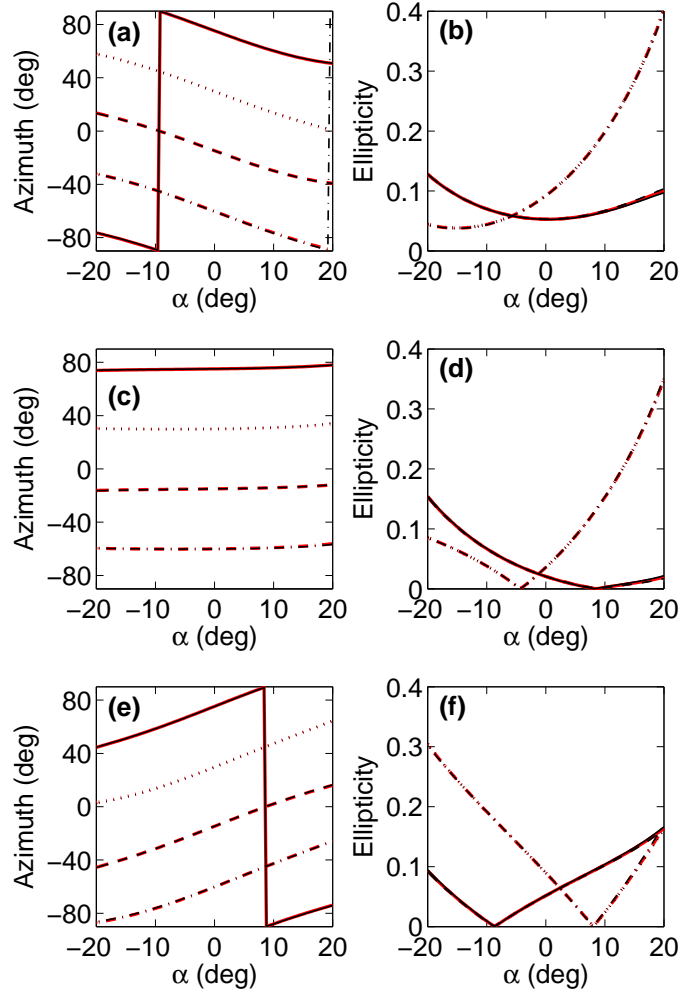}
\caption{\small Azimuth and ellipticity of output polarization states from a galvo scanner. (a) and (b):
$\beta = -20\deg$; (c) and (d): $\beta = 0\deg$; (e) and (f): $\beta = +20\deg$. Black lines refer to the output
states calculated using the Jones matrix of Eq.~\eqref{eqn:jones_galvo1}, while red lines have been obtained
assuming lossless mirrors. The polarization states of the input beam are $H$ (solid lines), $V$ (dashed lines),
+45 (dash-dotted lines) and -45 (dotted lines). In the right panels, dashed lines appear superimposed on solid lines,
while dotted lines appear superimposed on dash-dotted lines.}
\label{fig:states1}
\end{figure}

\begin{figure}[p]
\centering
\includegraphics[width=10cm]{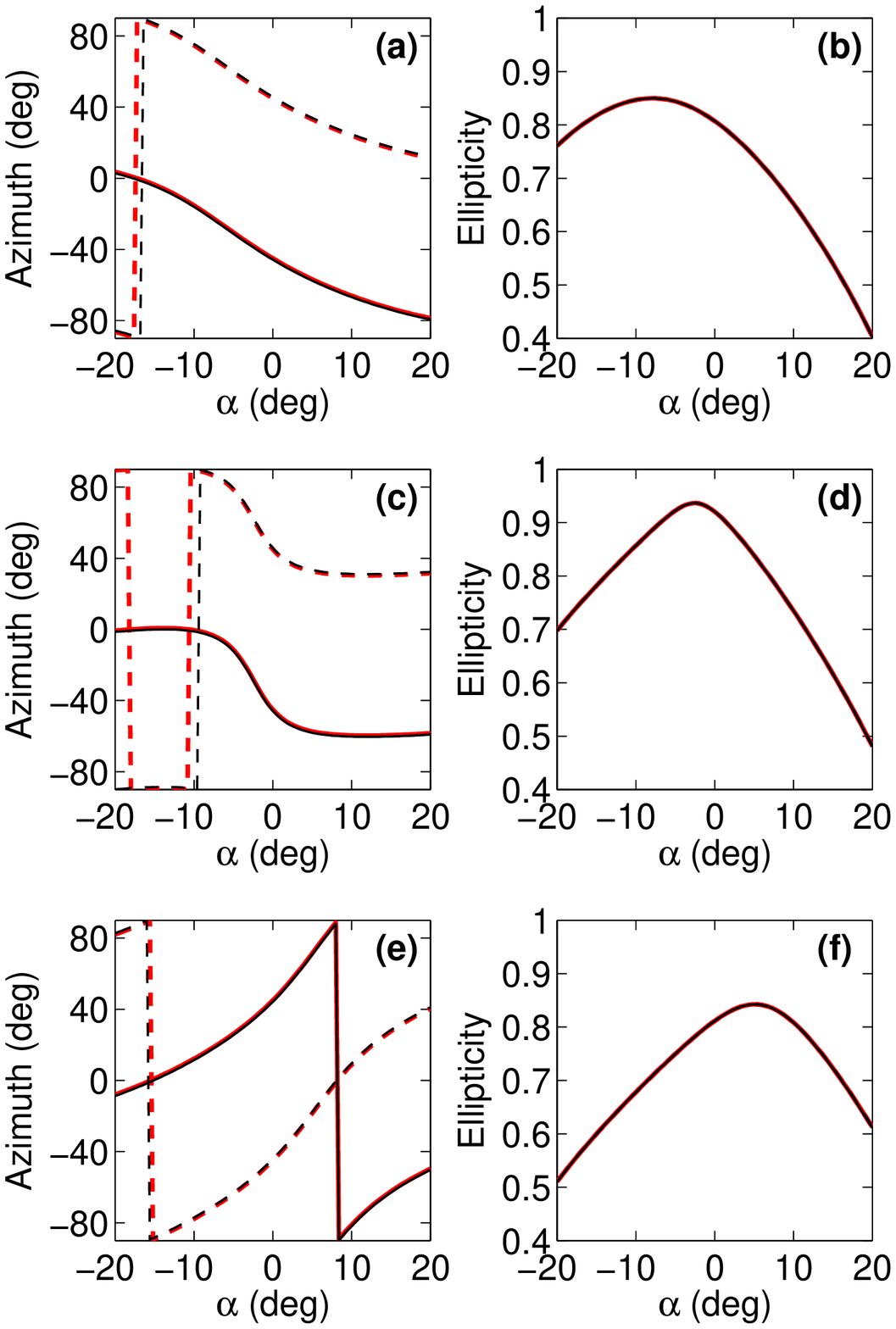}
\caption{\small Azimuth and ellipticity of output polarization states from a galvo scanner. (a) and (b):
$\beta = -20\deg$; (c) and (d): $\beta = 0\deg$; (e) and (f): $\beta = +20\deg$. Black lines refer to the output
states calculated using the Jones matrix of Eq.~\eqref{eqn:jones_galvo1}, while red lines have been obtained
assuming lossless mirrors. The polarization states of the input beam are $R$ (solid lines), and $L$ (dashed lines).
In the right panels, all lines appear superimposed.}
\label{fig:states2}
\end{figure}

As an useful example, we consider a galvo scanner with bare silver mirrors and a collimated laser beam at 850~nm.
Since both M1 and M2 are made of the same material, they also have the same complex refractive index $\tilde{n}
= 0.152 + 5.678 \, \ii$ at that wavelength \cite{solids}. We then take a number of polarization states of the input
beam and calculate the corresponding final state obtained by varying the rotation angles $\alpha$ and $\beta$ of
M1 and M2 in the range $[-20\deg, 20\deg]$ with steps of $0.1\deg$. The polarization states of the input beam are
chosen in order to fill as much as possible the Poincar\'{e} sphere, which means ellipticity in the range $[-1, 1]$
with steps of 0.01 and azimuth in the range $[-90\deg, 90\deg]$ with steps of $1.8\deg$.

For each initial state and for each mutual position of the galvo mirrors, we obtain the exact output state by
using the Jones matrix of Eq.~\eqref{eqn:jones_galvo1}, as well as the approximated version provided by
Eq.~\eqref{eqn:jones_galvo2}. In Fig.~\ref{fig:states1} we show the azimuth and the ellipticity of the output
states obtained considering four linear polarization states of the input beam (ellipticity equal to zero),
i.e. horizontal ($H$, $\mathbf{E}_1 = [1, 0]$), vertical ($V$, $\mathbf{E}_1 = [0, 1]$), linear at $45\deg$
(45, $\mathbf{E}_1 = [1, 1] / \sqrt{2}$), linear at $-45\deg$ (-45, $\mathbf{E}_1 = [1, -1] / \sqrt{2}$),
while in Fig.~\ref{fig:states2} we report the values obtained using right circular ($R$, $\mathbf{E}_1 =
[1, -\ii] / \sqrt{2}$) and left circular ($L$, $\mathbf{E}_1 = [1, \ii] / \sqrt{2}$) initial states
(ellipticity equal to one). Both Figures show the exact output states and the corresponding approximated
version, for comparison.

In order to define the degree of reliability of the lossless mirror approximation, we calculate the absolute
value of the difference between azimuth and ellipticity of the states obtained by using the two versions of
the Jones matrix. We find that the absolute value of the difference between the azimuth is typically lower
than $\sim 0.8 \deg$, while the absolute value of the difference between the ellipticity is typically lower
than $\sim 5 \times 10^{-3}$. For this reason, black and red curves related to the same initial polarization
state appear almost perfectly superimposed in Fig.~\ref{fig:states1} and \ref{fig:states2}. However, the
lossless mirror approximation cannot be used if the output polarization state is close to circular. Under this
condition and for particular combinations of the $\alpha$ and $\beta$ angle there might a large discrepancy
(even more than $10\deg$) between the azimuth of exact and approximated states.

\section{Rotation operators} \label{sec:rotop}

Following a notation commonly used in quantum mechanics, the approximated Jones matrix of a galvo scanner
[Eq.~\eqref{eqn:jones_galvo2}] can be represented as a product of rotation operators \cite{sakurai}:
\begin{equation}
\mathbf{M} \approx \exp(-\ii \sigma_y \eta_2) \, \exp(\ii \sigma_z \Phi_2) \, \exp(-\ii \sigma_y \eta_1) \,
\exp(\ii \sigma_z \Phi_1) \, \exp(-\ii \sigma_y \eta_0)~,
\end{equation}
where the Pauli matrices are
\begin{equation}
\sigma_x = \left[\begin{array}{cc}
0 & 1 \\
1 & 0
\end{array}\right]~, \;
\sigma_y = \left[\begin{array}{cc}
0 & -\ii \\
\ii & 0
\end{array}\right]~, \;
\sigma_z = \left[\begin{array}{cc}
1 & 0 \\
0 & -1
\end{array}\right]~.
\end{equation}
For this reason, the $\mathbf{M}$ matrix is unitary and unimodular.

\begin{figure}[p]
\centering
\includegraphics[width=10cm]{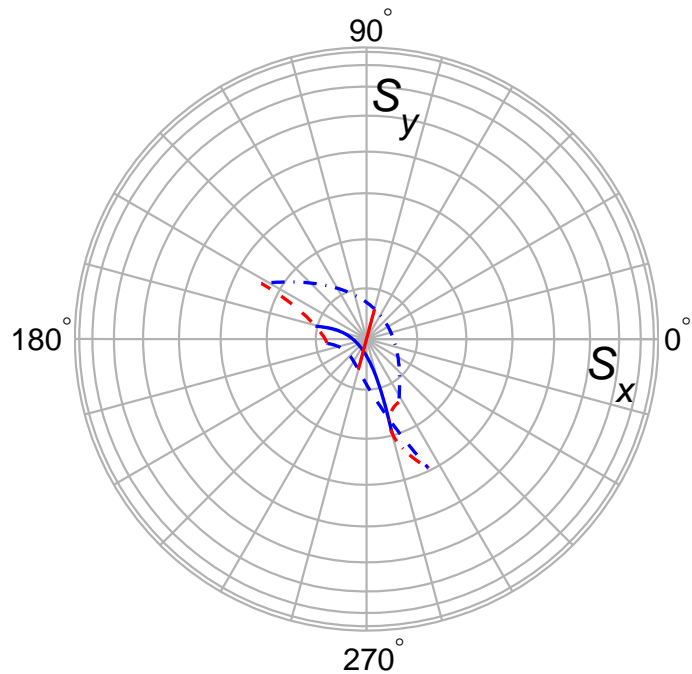}
\caption{\small Traces of the direction of the rotation axis associated to the Jones matrix of a galvo
scanner as a function of the $\alpha$ and $\beta$ angles. The lines are traced on the surface of the
Poincar\'{e} sphere, whose bottom part is depicted with a gray grid. Radial lines correspond to meridians drawn
at $15\deg$ intervals, while circles corresponds to parallels drawn at $10\deg$ intervals starting from the
south pole. Blue lines refer to a fixed $\beta$ ($\beta = -20\deg$ dashed line, $\beta = 0\deg$ solid line,
$\beta = 20\deg$ dash dotted) and $-20\deg \leq \alpha \leq 20\deg$, while red lines refer to a fixed $\alpha$
($\alpha = -20\deg$ dashed line, $\alpha = 0\deg$ solid line, $\alpha = 20\deg$ dash dotted) and
$-20\deg \leq \beta \leq 20\deg$.}
\label{fig:axis1}
\end{figure}

\begin{figure}[p]
\centering
\includegraphics[width=10cm]{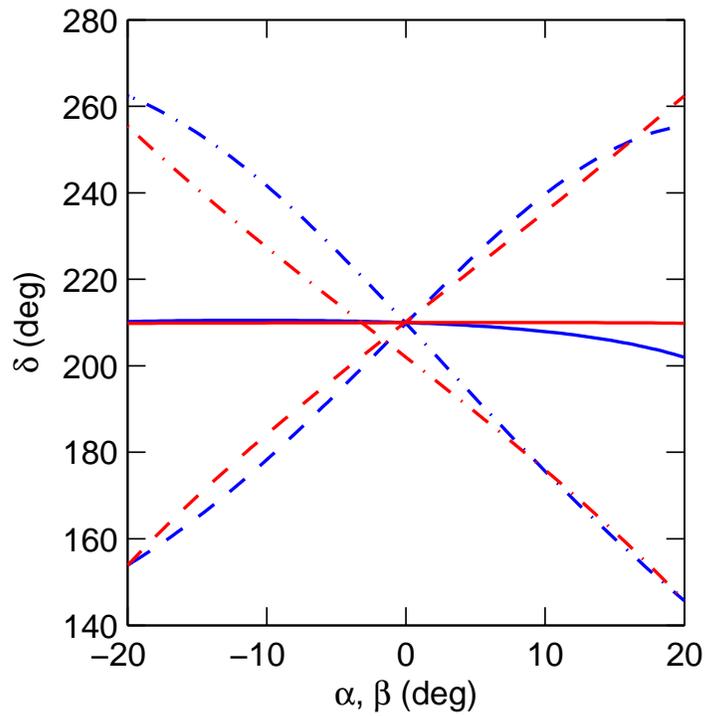}
\caption{\small Rotation angle $\delta$ associated to the Jones matrix of a galvo scanner as a function
of the $\alpha$ and $\beta$ angles. Blue lines refer to a fixed $\beta$ ($\beta = -20\deg$ dashed line,
$\beta = 0\deg$ solid line, $\beta = 20\deg$ dash dotted) and $-20\deg \leq \alpha \leq 20\deg$, while red lines
refer to a fixed $\alpha$ ($\alpha = -20\deg$ dashed line, $\alpha = 0\deg$ solid line, $\alpha = 20\deg$ dash
dotted) and $-20\deg \leq \beta \leq 20\deg$.}
\label{fig:axis2}
\end{figure}

It is well known that any unitary and unimodular matrix $\mathbf{U}$ represents a rotation on the Poincar\'{e} sphere:
\begin{equation} \label{eqn:rotation}
\mathbf{U} = \exp\left(-\ii \mathbf{n} \cdot \boldsymbol{\sigma} \frac{\delta}{2}\right)~,
\end{equation}
where $\boldsymbol{\sigma} = [\sigma_x, \sigma_y, \sigma_z]$, $\mathbf{n} = [\sin \vartheta \cos \varphi,
\sin \vartheta \sin \varphi, \cos \vartheta]$ is an unit vector defining the rotation axis ($\varphi$ is the longitude
and $\vartheta$ is the latitude) and $\delta$ is the rotation angle about $\mathbf{n}$. The $\mathbf{U}$
matrix can be always expressed in terms of a constant $a_0$ and a vector $\mathbf{a} = [a_x, a_y, a_z]$ as
\begin{equation}
\mathbf{U} = a_0 \mathbf{1} + \ii \mathbf{a} \cdot \boldsymbol{\sigma}=
\left[\begin{array}{cc}
a_0 + \ii a_z & \ii a_x + a_y\\
\ii a_x - a_y & a_0 - \ii a_z
\end{array}\right]~,
\end{equation}
where $\mathbf{1}$ is the identity matrix, $a_0 = \cos (\delta / 2)$, $|\mathbf{a}| = \sin (\delta / 2)$ and
$\mathbf{n} = \mathbf{a} / |\mathbf{a}|$. The transformation matrix of the galvo scanner expressed in
Eq.~\eqref{eqn:jones_galvo2} can be decomposed according to this scheme, thus giving:
\begin{equation} \label{eqn:rot_axis}
\left\{
\begin{aligned}
a_0 & = -\sin \Phi_1 \sin \Phi_2 \cos(\eta_0 - \eta_1 + \eta_2) + \cos \Phi_1 \cos \Phi_2 \cos(\eta_0 + \eta_1 + \eta_2) \\
a_x & = -\sin \Phi_1 \cos \Phi_2 \sin(\eta_0 - \eta_1 - \eta_2) - \cos \Phi_1 \sin \Phi_2 \sin(\eta_0 + \eta_1 - \eta_2) \\
a_y & = -\cos \Phi_1 \cos \Phi_2 \sin(\eta_0 + \eta_1 + \eta_2) + \sin \Phi_1 \sin \Phi_2 \sin(\eta_0 - \eta_1 + \eta_2) \\
a_z & = \sin \Phi_1 \cos \Phi_2 \cos(\eta_0 - \eta_1 - \eta_2) + \cos \Phi_1 \sin \Phi_2 \cos(\eta_0 + \eta_1 - \eta_2)
\end{aligned}
\right.~.
\end{equation}
For each configuration of the two mirrors of the galvo scanner, i.e. for each pair of rotation angles $\alpha$
and $\beta$, Eq.~\eqref{eqn:rot_axis} can be used to calculate the instantaneous rotation axis $\mathbf{n}$ and
the corresponding rotation angle $\delta$ that define the transformation of the polarization state of the input beam.

Figs.~\ref{fig:axis1} and \ref{fig:axis2} show the results obtained by considering a galvo scanner with the
characteristics described in Sect.~\ref{sec:lossless} for various combinations of $\alpha$ and $\beta$. Note that
the instantaneous rotation axis coincides with the negative $z$ axis for $\alpha = 0\deg$ and $\beta = 7.5\deg$.

\section{Conclusions}

We have studied the optical properties of a two-axis galvanometric optical scanner in order to understand how the
polarization state of an input beam changes at the output of the system. We obtained the transformation matrix that
maps the propagation direction of the input beam into the propagation direction of the output beam, as well as the
Jones matrix that maps the initial polarization state into the final one. Both these matrices have been expressed
in function of the rotation angles $\alpha$ and $\beta$, therefore permitting to predict the output polarization
state for any allowed position of two galvo mirrors. This change corresponds to a general rotation of the polarization
state in the Poincar\'{e} sphere, where both the instantaneous rotation axis and the rotation angle are both functions
of the $\alpha$ and $\beta$ angles.

Although the numerical results presented here has been obtained by considering a particular orientation of the two
galvo mirrors in their zero position, the analytical expressions can also be applied also to any other configuration
of the optical scanner simply by using the appropriate normal vectors, as well as to any kind of mirror having protection
coating by using the appropriate form of the Fresnel coefficients.

\section*{Acknowledgments}
The authors thank Valerio Pruneri and Juan P. Torres for helpful comments.

\end{document}